\documentclass[11pt,a4paper]{article}
\pdfoutput=1

\usepackage{jheppub}

\usepackage{bbold}
\usepackage{bm}
\usepackage{color}
\usepackage{dcolumn}
\usepackage{feynmf} 
\usepackage{graphics}
\usepackage{graphicx}
\usepackage{slashed}
\usepackage{tensor}
\usepackage{natbib}
\usepackage{hyperref}

\hypersetup{backref = true,pagebackref = true,hyperindex = true, colorlinks = true,
  breaklinks = true, urlcolor = blue, linkcolor = blue, bookmarks = true,
  bookmarksopen = true, citecolor=red}

\def\B{\langle }
\def\K{\rangle }

\def\Tr{\mbox{Tr}}

\def\al{\alpha}

\def\om{\omega}
\def\veps{\varepsilon}

\def\be{\begin{equation}}
\def\ee{\end{equation}}
\def\bea{\begin{eqnarray}}
\def\eea{\end{eqnarray}}
\def\bse{\begin{subequations}}
\def\ese{\end{subequations}}
\def\bc{\begin{center}}
\def\ec{\end{center}}

\def\ra{\rightarrow}

\def\nonum{\nonumber}

\def\I{{\rm i}}

\def\D{{\rm d}}
\def\Ord{{\rm O}}

\def\Sp{{\slashed p}}

\newcommand{\ie}{{\it i.e.}}
\newcommand{\eg}{{\it e.g.}}

\newcommand{\comment}[1]{}

\catcode`,\active

\catcode`\,12

\title{\boldmath Field theoretic renormalization study of interaction corrections to the universal ac conductivity of graphene}

\author[a]{S.~Teber}
\author[b]{and A.~V.~Kotikov}

\affiliation[a]{Sorbonne Universit\'e, CNRS, Laboratoire de Physique Th\'eorique et Hautes Energies, LPTHE, F-75005 Paris, France}
\affiliation[b]{Bogoliubov Laboratory of Theoretical Physics, Joint Institute for Nuclear Research, 141980 Dubna, Russia}

\emailAdd{teber@lpthe.jussieu.fr}
\emailAdd{kotikov@theor.jinr.ru}

\abstract{The two-loop interaction correction coefficient to the universal ac conductivity of disorder-free intrinsic graphene is computed with the help of a field
theoretic renormalization study using the Bogoliubov-Parasiuk-Hepp-Zimmermann prescription. Non-standard Ward identities imply that divergent subgraphs (related to Fermi velocity renormalization)
contribute to the renormalized optical conductivity. Proceeding either via density-density or via current-current correlation functions, a single well-defined
value is obtained: $\mathcal{C}= (19-6\pi)/12) = 0.01$ in agreement with the result first obtained by Mishchenko and which is compatible with experimental uncertainties.}

\begin{document} 
\maketitle
\flushbottom

\begin{fmffile}{fmfsigma}

\section{Introduction}

Transport properties of graphene and similar planar Dirac liquids have been the subject of extensive studies for more than a decade
now, see Ref.~\cite{Peres:2010mx} for a review. Of central interest for charge transport is the conductivity, $\sigma$,
which is in general a complicated function of, \eg, frequency ($\om$), momentum ($\vec q$\,), temperature ($T$), chemical potential ($\mu$) and scattering rates ($\Gamma$). 
A remarkable feature of the ideal case of an intrinsic ($\mu=0$) disorder-free graphene monolayer is that,
despite the vanishing density of states at the Fermi points, the relativistic-like nature of the charge carriers~\cite{PhysRev.71.622,Semenoff:1984dq} yields a non-zero universal 
	ac conductivity in the collisionless limit ($\om \gg T, \Gamma$):~\footnote{Throughout this paper, we work in units where $\hbar=k_\text{B}=1$.}
\be
\sigma_0 = \frac{e^2}{4}\, .
\label{sigma0}
\ee
This result, which was predicted long ago to hold for free Dirac fermions~\cite{Fradkin86.PhysRevB.33.3263,Lee93.PhysRevLett.71.1887,Ludwig94.PhysRevB.50.7526}, 
agrees to within $1$-$2\%$ with experiments in the optical regime, \eg, at $\om \sim 1$eV (visible range of the spectrum)~\cite{Nair:2008zz,PhysRevLett.101.196405}.
Still in the case of free fermions, adding other factors, \eg, $T$, $\mu$, $\Gamma$, ..., is rather non-trivial but leads to results~\cite{Gusynin:2006ym,Gusynin:2009-ac} 
which fit quite well the experimental data~\cite{Nair:2008zz,PhysRevLett.101.196405}. This is rather surprising because the long-range Coulomb interaction among charge carriers 
is not only unscreened but also supposed to be strong as witnessed by the fine structure constant of suspended graphene:
\be
\al_g = \frac{e^2}{4 \pi v} \approx 2.2 \, ,
\label{alg}
\ee
which is of the order of unity due to the fact that $v \approx c / 300$ where $v$ and $c$ are the Fermi and light velocities, respectively. 
Moreover, it is well known that Kohn's theorem~\cite{Kohn61PhysRev.123.1242} does not apply to pseudo-relativistic systems thereby allowing electron-electron interactions to affect Eq.~(\ref{sigma0}), 
see also the recent Ref.~\cite{Throckmorton:2018} and references therein for more.
 There has therefore been extensive theoretical works during the past decade devoted to understanding the intriguing effect of interactions on the 
optical conductivity of graphene in the collisionless limit, see, {\it e.g.}, Refs.~\cite{PhysRevB.83.195401,Herbut08.PhysRevLett.100.046403,Mishchenko2008,%
	Juricic:2010dm,Sheehy09.PhysRevB.80.193411,Abedinpour11.PhysRevB.84.045429,Teber:2012de,Sodemann12.PhysRevB.86.115408,%
	Kotikov:2013kcl,Gazzola13.0295-5075-104-2-27002,Rosenstein13.PhysRevLett.110.066602,Herbut+Mastropietro:PhysRevB.87.205445,%
	Teber:2014ita,Link16.PhysRevB.93.235447,Boyda:2016emg,Stauber17.PhysRevLett.118.266801,Teber:2018goo}, see also \cite{Teber:2016unz} for a short review. 
The latter can be defined via a density-density correlation function:
\be
\sigma(q_0) = - \lim_{\vec{q} \ra 0} \, \frac{\I q_0}{|\vec{q}\,|^2}\,\Pi^{00}(q_0,\vec{q}\,)\, ,
\label{sigma-dd}
\ee
where, in real time, $\Pi^{0 0} (t,\vec{q}\,) = \B T \rho(t,\vec{q}\,) \rho(0,-\vec{q}\,) \K$, $\rho$ is the charge density and $T$ the time-ordering operator. 
Equivalently, from current conservation, it can also be defined via a current-current correlation function (Kubo formula):
\be
\tilde{\sigma}(q_0) = \frac{1}{\I q_0}\,\frac{K^{11}(q_0,\vec{q}=0\,) + K^{22}(q_0,\vec{q}=0\,)}{2}\, ,
\label{sigma-cc}
\ee
where, in real time, $K^{i j} (t,\vec{q}\,) = \B T j^i(t,\vec{q}\,) j^j(0,-\vec{q}\,) \K$ and $\vec{j}$ is the charge current. 
In the case of weak short-range interactions, it was rigorously established that no interaction correction 
arises~\cite{PhysRevB.83.195401}.~\footnote{According to Ref.~\cite{Herbut+Mastropietro:PhysRevB.87.205445}, in the
case of weak short-range interactions, interaction corrections cancel out to
all orders in the renormalized expansion as shown in Ref.~\cite{PhysRevB.83.195401} as a
consequence of Ward identities and the irrelevance, in a renormalization
group sense, of the interaction (long-range interactions are, on the
other hand, only marginally irrelevant).}
 However, no exact result is available in the more realistic case of long-range interactions.	
Analytically, though interactions are strong, the problem is generally considered with the help of perturbation theory with a 
focus on the lowest order corrections to $\Pi^{00}(q)$ and $K^{ij}(q)$:
\begin{subequations}
\label{pert-exp}
\bea
	&&\sigma(q_0) = \sigma_0\, \bigg( 1 + \mathcal{C} \al_{g} + \Ord(\al_{g}^2) \bigg)\, , 
\\ 
	&&\tilde{\sigma}(q_0) = \sigma_0\, \bigg( 1 + \tilde{\mathcal{C}} \al_{g} + \Ord(\al_{g}^2) \bigg)\, , 
\eea
\end{subequations}
where the numbers $\mathcal{C}$ and $\tilde{\mathcal{C}}$ are the so-called first order interaction-correction coefficients.
On physical grounds, one expects that $\mathcal{C} = \tilde{\mathcal{C}}$, independent on the method used. 
It turns out that, despite the apparent simplicity of the task, different theoretical results have been found in the literature during the past ten years, see Tab.~\ref{tab:C-values} 
for a summary of some results. Following Refs.~\cite{Rosenstein13.PhysRevLett.110.066602,Teber:2014ita}, the main three values read:
\begin{subequations}
\label{c123}
\bea
\mathcal{C}^{(1)}  = \frac{25-6\pi}{12} \approx 0.512\, ,
\label{c1}\\
\mathcal{C}^{(2)}  = \frac{19-6\pi}{12} \approx 0.013\, ,
\label{c2}\\
\mathcal{C}^{(3)}  = \frac{11-3\pi}{6} \approx 0.263\, .
\label{c3}
\eea
\end{subequations}
As can be seen from Tab.~\ref{tab:C-values}, the most commonly accepted result is, up to date, the value $\mathcal{C}^{(2)}$ of Mishchenko~\cite{Mishchenko2008} since it has been recovered 
by a majority of groups. Incidentally, this is also the only result, among those of Eqs.~(\ref{c123}), which is consistent with the experimental 
uncertainties~\cite{Nair:2008zz,PhysRevLett.101.196405} as $\mathcal{C}^{(2)} \al_g \approx 2\%$.

\begin{center}
\renewcommand{\tabcolsep}{0.25cm}
\renewcommand{\arraystretch}{1.5}
\begin{table}
\begin{tabular}{|c|c|c|}
\hline
             $\mathcal{C}$     		                        &            {\bf Method}                                          &       {\bf Year} \\
\hline
\hline
$\mathcal{C}^{(1)}  = \frac{25-6\pi}{12} \approx 0.512$       	& Eq.~(\ref{sigma-cc}) hard cut-off			   &    2008~\cite{Herbut08.PhysRevLett.100.046403}    \\
\hline
\hline
\textcolor{red}{$\mathcal{C}^{(2)}  = \frac{19-6\pi}{12} \approx 0.013$}		& Eq.~(\ref{sigma-dd}) hard cut-off	   &    2008~\cite{Mishchenko2008} \\
\cline{2-3}
			     					& Eq.~(\ref{sigma-cc}) and kinetic equations, soft cut-off   &    2008~\cite{Mishchenko2008}           \\
\cline{2-3}
        							& Eqs.~(\ref{sigma-cc}) hard cut-off			   &    2009~\cite{Sheehy09.PhysRevB.80.193411}            \\
\cline{2-3}
        							& Eq.~(\ref{sigma-dd}) hard cut-off			   &    2011~\cite{Abedinpour11.PhysRevB.84.045429}        \\
\cline{2-3}
        							& Eq.~(\ref{sigma-dd}) hard cut-off			   &    2012~\cite{Sodemann12.PhysRevB.86.115408}           \\
\cline{2-3}
                                                                & Eqs.~(\ref{sigma-cc}) hard cut-off, implicit regularization  &  2013~\cite{Gazzola13.0295-5075-104-2-27002}           \\
\cline{2-3}
                                                                & Eqs.~(\ref{sigma-dd}), (\ref{sigma-cc}),  DR + CR   &    2014~\cite{Teber:2014ita}            \\
\cline{2-3}
                                                                & lattice (tight-binding) simulations 				   &    2016~\cite{Link16.PhysRevB.93.235447}           \\
\cline{2-3}
                                                                & Quantum Monte Carlo calculations				   &    2016~\cite{Boyda:2016emg}           \\
\cline{2-3}
($0.05$)     	                                                & Hartree-Fock simulations (self-screened)                         &    2017~\cite{Stauber17.PhysRevLett.118.266801}\\
\cline{2-3}
                                                                & \textcolor{red}{Eqs.~(\ref{sigma-dd}), (\ref{sigma-cc}),  DR + BPHZ}  &    \textcolor{red}{2018}            \\
\hline
\hline
$\mathcal{C}^{(3)}  = \frac{11-3\pi}{6} \approx 0.263$		& kinetic equations, hard cut-off				   &    2008~\cite{Mishchenko2008}                  \\
\cline{2-3}
  			                                        & Eqs.~(\ref{sigma-dd}), (\ref{sigma-cc}), DR          &  2010~\cite{Juricic:2010dm}          \\
\cline{2-3}
								& lattice (tight-binding) simulations				   &    2013~\cite{Rosenstein13.PhysRevLett.110.066602}\\
\cline{2-3}
($1/4=0.25$)                                                    & Hartree-Fock simulations (unscreened)                            &    2017~\cite{Stauber17.PhysRevLett.118.266801}\\
\hline
\hline
$\mathcal{C}^{*}  = \frac{92-9\pi^2}{18\pi} \approx 0.056$          & Eq.~(\ref{sigma-cc}) DR + CR ($v/c \ra 1$)	           &  2012~\cite{Teber:2012de} 2013~\cite{Kotikov:2013kcl}                  \\
								& Eq.~(\ref{sigma-cc}) DR + BPHZ ($v/c \ra 1$)	           &  2017~\cite{Teber:2016unz} 2018~\cite{Teber:2018goo}		\\
\hline
\end{tabular}
    \caption{Some values of $\mathcal{C}$ obtained over the years together with elements of the different methods used. 
	In case of numerical simulations, we cite the numerical value obtained whenever available and when it slightly differs from the main 
	three results found in the literature, $\mathcal{C}^{(i)}$ ($i=1,2,3$), Eqs.~(\ref{c123}). 
	For the sake of completeness, the value at the IR fixed point, $\mathcal{C}^{*}$, has been also added. 
	DR is for dimensional regularization. CR is for conventional renormalization. The result derived in this paper appears in red.}
    \label{tab:C-values}
\end{table}
\end{center}

At this point, we note that there is a limit where the result for the interaction correction coefficient does not raise any doubt: the deep infra-red (IR) limit corresponding 
to the Lorentz-invariant fixed point~\cite{Gonzalez:1993uz}. The later arises from the long-range Coulomb interaction among the Dirac fermions which
enforces the flow of the Fermi velocity, $v \approx c/300$ at the present experimentally accessible scales, to the velocity of light, $c$, in the IR with a corresponding flow of
the fine structure constant, $\al_g \approx 2.2$, to the usual fine structure constant, $\al \approx 1/137$, in the IR. The IR fixed point therefore corresponds to an
ultra-relativistic limit with fully-retarded interactions ($v/c \ra 1$). The corresponding conductivity reads~\cite{Teber:2012de,Kotikov:2013kcl,Teber:2018goo,Teber:2016unz}:
\be
\tilde{\sigma}(q_0) = \sigma_0\, \bigg( 1 + \mathcal{C}^* \al + \Ord(\al^2) \bigg)\, , \qquad \mathcal{C}^* = \frac{92-9\pi^2}{18\pi}\, .
\label{sigma-cc*}
\ee
Of course, at the fixed point $\al = 1/137$ and the product $\mathcal{C}^* \al \approx 10^{-4}$ is very small leading to almost unobservable effects. However, it is interesting to note that
$\mathcal{C}^{*} = 0.056$, a value which is of the same order of magnitude as $\mathcal{C}^{(2)}$.~\footnote{The interest in comparing $\mathcal{C}^{(2)}$ to $\mathcal{C}^{*}$ 
(rather than $\mathcal{C}^{(2)}\alpha_g$ to $\mathcal{C}^{*}\alpha$) comes from the fact that there is a more general model (model I in the terminology of Ref.~\cite{Teber:2018goo}) 
describing interactions in graphene which is valid for arbitrary $v/c$. So there is actually a non-trivial interaction correction {\it function}
$\mathcal{C}(v/c)$ which encodes relativistic corrections (the dependence of the fine structure constant on $v$ is trivial). Presently, only the limiting
values $\mathcal{C}(v/c \rightarrow 0) = \mathcal{C}^{(2)}$ (see the proof in the following pages) and $\mathcal{C}(v/c \rightarrow 1) = \mathcal{C}^{*}$ are
known. It is surprising that these values are of the same order as if $\mathcal{C}(v/c)$ was only weakly dependent on $v/c$. A study of $\mathcal{C}(v/c)$ for arbitrary
$v/c$ is beyond the scope of the present paper and we leave it for our future investigations.} The result of Eq.~(\ref{sigma-cc*}) was derived in 
Refs.~\cite{Teber:2012de,Kotikov:2013kcl} with the help of multi-loop techniques together with conventional renormalization. 
In Ref.~\cite{Teber:2018goo}, see also the review \cite{Teber:2016unz}, the 
Bogoliubov-Parasiuk-Hepp-Zimmermann (BPHZ), see Refs.~\cite{Bogoliubov:1957gp,Hepp:1966eg,Zimmermann1969} as well as Ref.~\cite{Collins:1984xc} for a textbook, 
prescription was used.~\footnote{Let's also note the more recent Hopf algebraic formulation~\cite{Kreimer:1997dp} of renormalization (see Ref.~\cite{Panzer:2014kia} for a
recent review). Its application to our model is beyond the scope of our present study.} 
This is a powerful renormalization prescription, which allows to systematically construct counter-terms on a diagram by diagram basis and can be generalized to higher orders.
It was then found in Refs.~\cite{Teber:2018goo,Teber:2016unz} 
that, at the fixed point, divergent subgraphs (due to wave function renormalization) 
cancel each other due to standard, \ie, similar to those appearing in relativistic quantum electrodynamics (QED), Ward identities yielding the fixed point value of Eq.~(\ref{sigma-cc*}). 
Let's also remark that at the fixed point $\al \ll 1$ and the result (\ref{sigma-cc*}) is reliable. But clearly, at higher energy scales such as those relevant to experiments, results based on the loop 
expansion (\ref{pert-exp}) may receive strong corrections from higher orders and the perturbative approach is rightly questionable. 
A way to overcome this difficulty may be to reorder the perturbative series in the form of a $1/N$-expansion or, in other words, to perform a random phase approximation-like resummation, see, {\eg},
Ref.~\cite{PhysRevLett.113.105502} for an attempt to carry out next-to-leading order (NLO) computations. 
This is a very interesting suggestion which certainly requires further study. In the following, we shall pursue a more modest goal
and restrict ourselves to the study of Eq.~(\ref{pert-exp}) without any additional resummation. The reason is that, in our opinion, the problem of an {\it accurate} evaluation of NLO
corrections cannot reasonably be addressed before a {\it full} understanding of the first few orders of the loop expansion is achieved.

In the present paper, we reconsider the computation of the first order interaction correction to the minimal conductivity of graphene in the standard but more subtle non-relativistic 
limit with instantaneous interactions ($v/c \ra 0$). Our analysis clarifies the fact that the origin of the discrepancy between the different results
found in this limit, the $\mathcal{C}^{(i)}$ ($i=1,2,3$) of Eqs.~(\ref{c123}), does not lie in the regularization method but, more simply, in the renormalization procedure itself. This was first revealed 
in Ref.~\cite{Teber:2014ita} with a proof obtained with the help of conventional renormalization. 
In the following, we will give a stronger proof based on the BPHZ renormalization prescription. In particular, we will explicitly show that, in the present non-relativistic case, 
non-standard Ward identities imply that divergent subgraphs (related to Fermi velocity renormalization) do contribute to the renormalized optical conductivity.
Proceeding either via density-density or current-current correlation functions and properly taking into account of these counter-terms will provide a clear explanation 
for why radiative corrections to the optical conductivity of graphene are finite and perfectly well determined.~\footnote{We are therefore
not in the field theoretic situation reviewed by Jackiw in Ref.~\cite{Jackiw:1999qq}.}
Anticipating the conclusion, we will show that our approach is in favour of the result first derived by Mishchenko~\cite{Mishchenko2008}. Our final result reads:
\be
\mathcal{C}_r = \tilde{\mathcal{C}}_r = \mathcal{C} + \mathcal{C'} = \mathcal{C}^{(2)}, \qquad \mathcal{C} = \mathcal{C}^{(3)}, \quad \mathcal{C'} = -\frac{1}{4}\, ,
\label{c-DR}
\ee
where $r$ stands for renormalized and $\mathcal{C'}$ originates from one-loop counterterms.

The paper is organized as follows. In Sec.~\ref{Sec:model}, we introduce the basic model, the corresponding Feynman rules and the notations that will be used throughout the text.
In Sec.~\ref{Sec:OneLoop}, we briefly study the model at one-loop (self-energies and Ward identities) introducing the basic ingredients required for the two-loop computation. 
In Sec.~\ref{Sec:TwoLoop}, we focus on the optical conductivity at two-loop and compute the counter-terms with the help of the BPHZ prescription.
In Sec.~\ref{Sec:Conclusion} we summarize our results and conclude. Some useful integrals are provided in App.~\ref{App:MI} and
App.~\ref{App:Two-loop} contains some additional illustration of the singular subgraphs in two-loop diagrams.

\section{Model, Feynman rules and renormalization} 
\label{Sec:model}

The effective low-energy action (model II in the terminology of Ref.~\cite{Teber:2018goo}) that we wish to consider reads:
\bea
S = \int \D t\, \D^{2} x\, \bar{\psi}_\sigma \left[ \gamma^0 \left( \I \partial_t -eA_0 \right) + \I v \vec{\gamma} \cdot \vec{\nabla}\,\right] \psi^\sigma 
+ \frac{1}{2}\,\int \D t\, \D^{3} x\, \left( \vec{\nabla} A_0 \right)^2 \, ,
\label{model-inst}
\eea
where $v$ and $e$ are the bare Fermi velocity and charge, respectively,
$\psi_\sigma$ is a four-component spinor field describing a fermion of specie $\sigma$ ($\sigma=1,\cdots,N_F$ and for graphene $N_F=2$) and
$A_0$ is the gauge field mediating the instantaneous Coulomb interaction.
The Dirac matrices, $\gamma^\mu = (\gamma^0,\vec{\gamma}\,)$ satisfy the usual algebra $\{\gamma^\mu,\gamma^\nu \} = 2g^{\mu \nu}$,
with metric tensor $g^{\mu \nu} = \text{diag}(+,-,-)$. 

From Eq.~(\ref{model-inst}), the bare momentum space fermion propagator reads:
\be
S_0(p) = \frac{i\Sp}{p^2}\, , \qquad \Sp = \gamma^\mu p_\mu = \gamma^0 p_0 - v \vec{\gamma}\cdot \vec{p}\, ,
\label{fermion-prop0}
\ee
where we use pseudo-relativistic notations with $p^\mu=(p^0,v \vec{p}\,)$.
Because of the absence of any retardation effect (non-relativistic limit corresponding to $v/c \ra 0$), the effective photon propagator reduces to the instantaneous Coulomb interaction and reads:
\be
V_0(\vec{q}\,) = \frac{\I}{2 (|\vec{q}\,|^2)^{1/2}}\, .
\label{gauge-field-prop0}
\ee
Moreover, vector photons decouple and the bare vertex reduces to the temporal part: 
\be
-\I e \Gamma_0^0 = -\I e \gamma^0 \, .
\label{vertex0}
\ee
For our future purposes, it will be nevertheless convenient to define $\Gamma^\mu = (\Gamma^0, \vec{\Gamma}\,)$ such that $\Gamma_{0}^\mu = \gamma^\mu = (\gamma^0, \vec{\gamma}\,)$
and employ the following graphical notations:
\begin{subequations}
\label{Gamma_mu}
\bea
&&-\I e \gamma^0 = ~~ 
      \parbox{20mm}{
        \begin{fmfgraph*}(20,10)
          \fmfleft{p}
          \fmfright{ei,eo}
          \fmf{boson}{p,v}
          \fmf{vanilla}{ei,v}
          \fmf{fermion}{v,eo}
          \fmfv{decor.shape=circle,decor.filled=empty,decor.size=2thick}{v}
        \end{fmfgraph*}
      } ~~, 
\\
&&-\I e \vec{\gamma} = ~~
      \parbox{20mm}{
        \begin{fmfgraph*}(20,10)
          \fmfleft{p}
          \fmfright{ei,eo}
          \fmf{boson}{p,v}
          \fmf{vanilla}{ei,v}
          \fmf{fermion}{v,eo}
	\fmfv{decor.shape=circle,decor.filled=shaded,decor.size=2thick}{v}
        \end{fmfgraph*}
      } ~~, 
\\
&&-\I e \gamma^\mu = ~~
      \parbox{20mm}{
        \begin{fmfgraph*}(20,10)
          \fmfleft{p}
          \fmfright{ei,eo}
          \fmf{boson}{p,v}
          \fmf{vanilla}{ei,v}
          \fmf{fermion}{v,eo}
          \fmfdot{v}
        \end{fmfgraph*}
      } ~~.
\eea
\end{subequations}

In the following, we will use conventional dimensional regularization~\cite{Kilgore:2011ta}. The above Feynman rules stay the same but momenta, 
Dirac matrices and metric tensor are extended to span a $D_e$-dimensional space (keeping $\Tr \left[ {\bf 1} \right] = 4 N_F$) with 
\be
D_e = 2 - 2\veps_\gamma\, .
\label{def:De}
\ee
All bare parameters and fields are then related to renormalized ones via renormalization constants in a standard way: 
%
\begin{subequations}
\label{renormZ}
\bea
	&&\psi = Z_{\psi}^{1/2} \psi_r, \qquad A_0 = Z_{A_0}^{1/2} A_{0r}, 
	\\
	&&e = Z_e e_r \mu^{\veps_\gamma}, \qquad v = Z_v v_r\, ,
\eea
\end{subequations}
where:
\be
Z_e = \frac{Z_{\Gamma^0}}{Z_\psi Z_{A_0}^{1/2}}\, ,
\ee
and $\mu$ is the so-called renormalization scale in the minimal subtraction ($\text{MS}$) scheme. The latter
is related to the corresponding parameter $\overline{\mu}$ in the modified minimal subtraction ($\overline{\text{MS}}$) scheme with the help of:
\be
\overline{\mu}^{\,2} = 4\pi e^{-\gamma_E} \mu^2\, ,
\label{muMSbar}
\ee
where $\gamma_E$ is Euler's constant. The attractive feature of the $\text{MS}$ scheme is that the renormalization constants take the simple form:
\be
Z_x(\al_{gr}) = 1 + \delta Z_x (\al_{gr}) = 1 + \sum_{l=1}^\infty \sum_{j=1}^l Z_x^{(l,j)}\,\frac{\al_{gr}^l}{\veps_\gamma^j}\, ,
\label{Zx}
\ee
where $x \in \{ \psi,A,e,v,\Gamma^\mu \}$, $\al_{gr} = e_r^2/(4\pi v_r)$ is the renormalized coupling constant and
$l$ runs over the number of loops at which ultra-violet (UV) singularities are subtracted. The $Z_x$ do not depend on momentum; furthermore,
the dependence on $\mu$ is only through $\al_{gr}$. So the $Z_x$ depend only on $\al_{gr}(\mu)$ and $\veps$.
The basic correlation functions of the model are renormalized as follows:
\begin{subequations}
\bea
	&&S(k) = Z_\psi S_r(k)\, , 
	\\
	&&V(\vec{q}\,) = Z_{A_0} V_{r}(\vec{q}\,)\, , 
	\\
	&&\Gamma^\mu(k,q) = Z_{\Gamma^\mu}^{-1} \Gamma_r^\mu(k,q)\, ,
\eea
\end{subequations}
where the notation implicitly assumes that, besides the $Z_x$, only the renormalized quantities depend on $\mu$.

For the model of Eq.~(\ref{model-inst}), once integrated over the third spacial direction the effective action of the free gauge field becomes non-local.
This implies that the gauge field is not renormalized: $Z_{A_0}=1$~\cite{Vasil'evbook,PhysRevLett.80.5409,PhysRevLett.87.137004}.
As a consequence, the charge is not renormalized: $Z_e=1$. The renormalization of the coupling constant is therefore entirely due to the renormalization of the velocity which is the only
running parameter of the model:
\be
\al_g = Z_{\al} \,\al_{gr}, \qquad Z_{\al} = Z_v^{-1}\, .
\label{alr}
\ee
The velocity $\beta$-function is then defined as:
\begin{subequations}
\bea
	&&\beta_v(\al_{gr}) = \mu \frac{\partial v_{r}}{\partial \mu}\bigg|_B = \sum_{l=0}^\infty \beta_{v,l} \al_{gr}^{l+1}\, , 
	\\ 
	&&\beta_{v,l} = 2 v_r\,(l+1)\,Z_v^{(l+1,1)}\, ,
\label{gm:def:betav}
\eea
\end{subequations}
where the subscript $B$ indicates that bare parameters, which do not depend on $\mu$, are fixed.

\section{One-loop analysis}
	\label{Sec:OneLoop}

\begin{figure}
        \includegraphics{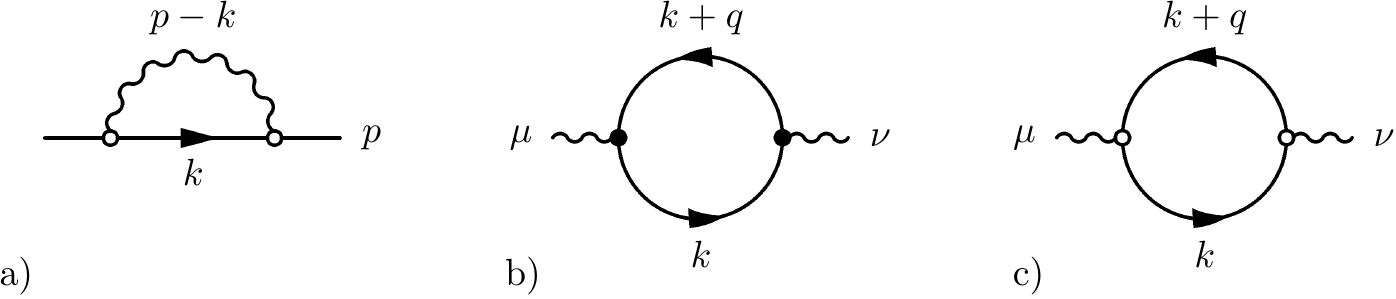}
  \caption{\label{fig:one-loop}
  One-loop: a) fermion self-energy and photon self-energies: b) $\Pi^{\mu \nu}$, c) $\Pi^{00}$.}
\end{figure}

We start by analyzing model of Eq.~(\ref{model-inst}) at one-loop. 

\subsection{One-loop fermion and photon self-energies}

Let's first recall some elementary results for clarity, see Ref.~\cite{Teber:2014ita} for more details. The one-loop fermion self-energy, Fig.~\ref{fig:one-loop}a, is defined as:
\be
-i\Sigma_1(k) = \int [\D^{d_e} q]\, (-\I e \gamma^0) \,S_0(k+q)\, (-\I e \gamma^0)\,V_0(q)\, ,
\label{def:fermion-self}
\ee
where $d_e = 1 + D_e$ is the space-time dimension. The following parametrization is useful:
\be
\Sigma(k) = \gamma^0 k_0 \,\Sigma_\om(k^2) - v \vec{\gamma} \cdot \vec{k}\,\,\Sigma_k(k^2)\, ,
\label{gm:Sigma:param}
\ee
which is such that:
\be
\Sigma_{\om}(k^2) = \frac{\Tr[\gamma^0 k_0\,\Sigma(k)]}{4N_F k_0^2}\, , \quad
\Sigma_{k}(k^2) = \frac{\Tr[\vec{\gamma} \cdot \vec{k}\,\Sigma(k)]}{4N_F v |\vec{k}\,|^2}\, .
\label{fsigma-param}
\ee
Using this parametrization together with the standard rules for integrating massless Feynman diagrams straightforwardly yields:
%
\begin{flalign}
&\Sigma_{1k}(|\vec{k}\,|^2) =  -\frac{\al_{gr}}{8}\, {\left(\frac{\overline{\mu}^{\,2}}{|\vec{k}\,|^2}\right)}^{\veps_\gamma} \, 
e^{2\gamma_E\veps_\gamma}\,G(1/2,1/2)\,  ,
\label{fsigma2-ka} 
\end{flalign}
%
where $G(\alpha,\beta)$ is defined in Eq.(\ref{massless-p-integral}) and $\Sigma_{1\om}(k^2) =0$.
 The factor $G(1/2,1/2)$ contains the pole $1/\veps_\gamma$ at $\veps_\gamma \to 0$ and is such that:
%
\begin{flalign}
G(1/2,1/2)=  \Gamma(1+\veps_\gamma) \, \left( \frac{1}{\veps_\gamma} + 4 \log 2 + \Ord(\veps_\gamma) \right)\,  .
\label{Gexp} 
\end{flalign}
%
Performing 
the $\veps_\gamma$ expansion~\cite{Teber:2014ita} then yields: 
%
\begin{flalign}
\Sigma_{1k}(|\vec{k}\,|^2) =  -\frac{\al_{gr}}{8}\, \left( \frac{1}{\veps_\gamma} - L_k + 4 \log 2 + \Ord(\veps_\gamma) \right)\,  ,
\label{fsigma2-k} 
\end{flalign}
%
where $L_k = \log(|\vec{k}\,|^2/\overline{\mu}^{\,2})$. From the above results, we see that there is a momentum but no frequency-dependence of 
the one-loop fermion self-energy due the instantaneous nature of the interaction. 
At one-loop, there is therefore a Fermi velocity renormalization~\cite{Gonzalez:1993uz} but no wave-function renormalization: 
\begin{subequations}
\label{Zv+Zpsi}
\bea
	&&Z_{1v} = 1 + \delta Z_{1v}, \qquad \delta Z_{1v}= -\frac{\al_{gr}}{8 \veps_\gamma}\, ,
\label{Zv} \\
	&&Z_{1\psi} = 1 + \delta Z_{1\psi}, \qquad \delta Z_{1\psi} = 0\, .
\label{Zpsi}
\eea
\end{subequations}
From Eq.~(\ref{gm:def:betav}) we see that the velocity beta-function is negative: 
\be
\beta_v = - v_r \al_{gr}/4 + \Ord(\al_{gr}^2)\, ,
\ee
implying that Fermi velocity grows in the infrared~\cite{Gonzalez:1993uz}.
Graphically, these results can be summarized as follows:
\be
\delta Z_{1v} = \mathcal{K}\bigg[ \Sigma_{1k}(k^2) \bigg] ~ = ~~
\mathcal{K}\bigg[~~
      \parbox{15mm}{
    \begin{fmfgraph*}(15,15)
      \fmfleft{in}
      \fmfright{out}
      \fmf{plain}{in,vi}
      \fmf{plain,tension=0.2}{vi,vo}
      \fmf{boson,left,tension=0.2}{vi,vo}
      \fmf{plain}{vo,out}
      \fmfv{decor.shape=circle,decor.filled=empty,decor.size=2thick}{vi,vo}
    \end{fmfgraph*}
}~~
\bigg] ~ = ~ - \frac{\al_{gr}}{8 \veps_\gamma}\, ,
\label{dZ1v}
\ee
where the pseudo Lorentz structure of the diagram in the brackets has been projected out, \ie, the displayed graph corresponds to $\Sigma_{1k}(k^2)$ and not $\Sigma_{1}(k)$, and 
the $\mathcal{K}$ operator is defined as:
\be
\mathcal{K}~\left( \sum_{n=-\infty}^{+\infty} \frac{c_n}{\veps_\gamma^n} \right) = \sum_{n=1}^{+\infty} \frac{c_n}{\veps_\gamma^n}\, .
\label{def:Sing}
\ee
Similar notations and conventions will be used in the following.

The one-loop photon self-energy, Fig.~\ref{fig:one-loop}b, is defined in the usual way as:
\be
\I \Pi_1^{\mu \nu}(q) = - \int [\D^{d_e}k]\, \Tr \left[ (-\I e \gamma^\mu)\,S_0(k+q)\,(-\I e \gamma^\nu)\,S_0(k) \right] \, .
\label{pi1munu}
\ee
Focusing on $\Pi^{00}$, performing the integrations and the $\veps_\gamma$ expansion~\cite{Teber:2014ita} yields:
\be
\Pi_1^{00}(q_{0},\vec{q} \ra 0) = -\frac{N_F e_r^2}{8}\,\frac{|\vec{q}\,|^2}{\I q_0}\,\biggl( 1 - (1 + L_{q_0})\,\veps_\gamma + \Ord(\veps_\gamma^2) \biggr)\, ,
\label{pi100}
\ee
which is finite as expected and where $L_{q_0}=\log(-q_0^2/(4v^2 \overline{\mu}^{\,2}))$.
Combining Eqs.~(\ref{sigma-dd}) and (\ref{pi100}), we arrive at the (renormalized) one-loop conductivity: 
\be
\sigma_{1r}(q_0) = \frac{N_F e_r^2}{8}\, , 
\label{sigma1}
\ee
which, for $N_F=2$, corresponds to the well known universal ac conductivity $\sigma_0$, Eq.~(\ref{sigma0}).

We may proceed in a similar way with the help of the Kubo formula Eq.~(\ref{sigma-cc}). 
The transversality of $\Pi^{\mu \nu}$ allows to parametrize it as follows:
\be
\Pi^{\mu \nu}(q) = (g^{\mu \nu}q^2 - q^\mu q^\nu)\,\Pi(q^2), \quad
\Pi(q^2) = \frac{- \Pi^\mu_\mu(q)}{(d_e-1)(-q^2)}\, .
\label{def:pi2}
\ee
Then:
\be
\tilde{\sigma}(q_0) = \I q_0 \, K(q_0), \qquad K(q_0) = v^2 \Pi(q_0^2,|\vec{q}\,|^2 \ra 0) \, .
\label{sigma-cc2}
\ee
Performing the integrations and the $\veps_\gamma$ expansion~\cite{Teber:2014ita} yields:
\bea
K_1(q_{0}) = \,\frac{N_F \, e_r^2}{8 \,\I q_0}\,\biggl( 1 - (1 + L_{q_0})\,\veps_\gamma + \Ord(\veps_\gamma^2) \biggr)\, . 
\label{K1-res}
\eea
Combining Eqs.~(\ref{sigma-cc2}) and (\ref{K1-res}), we arrive once again at: 
\be
\tilde{\sigma}_{1r}(q_0) = \frac{N_F e_r^2}{8}\, , 
\label{sigmat1}
\ee
in perfect agreement with Eq.~(\ref{sigma1}).

\begin{figure}
        \includegraphics{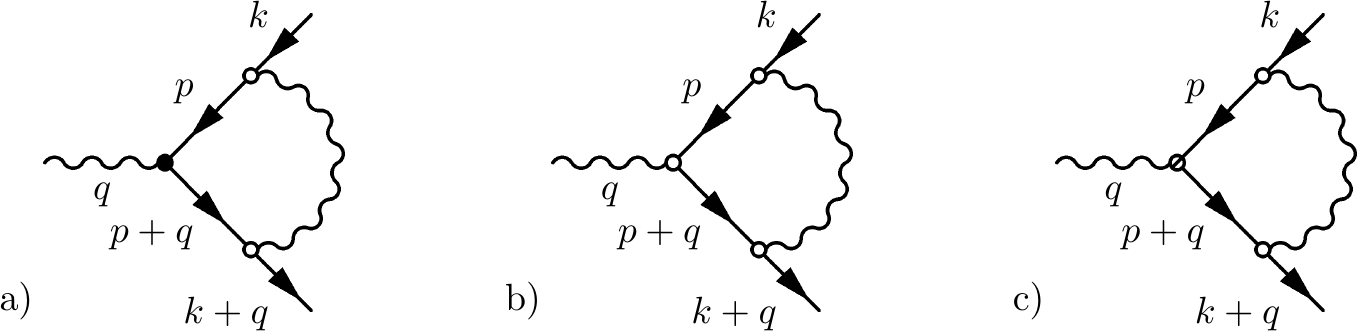}
  \caption{\label{fig:one-loop-vertex}
  One-loop vertex parts: a) $\Lambda_1^\mu$, b) $\Lambda_1^0$ and c) $\vec{\Lambda}_1$.}
\end{figure}

\subsection{One-loop fermion-photon vertex and Ward identities}

At this point it is also instructive to look at the vertex part: $\Gamma^\mu = \gamma^\mu + \Lambda_1^\mu + \Ord(\al_g^2)$ where $\Lambda_1^\mu$ is the 
one-loop correction, see Fig.~\ref{fig:one-loop-vertex}a. The latter is defined as:
\begin{flalign}
-\I e \Lambda_1^\mu(k,q) = \int [\D^{d_e}p]\, V_0(p-k)\,(-\I e\gamma^0)\,S_0(p+q)(-ie\gamma^\mu)\,S_0(p)(-\I e \gamma^0)\, .
\label{def:vertex}
\end{flalign}
We first consider the temporal part, Fig.~\ref{fig:one-loop-vertex}b. In order to single out it's UV divergent part we will evaluate it for $k=q=0$. 
Performing the trace and going to euclidean space ($q_0=\I q_{E0}$) yields:
\begin{flalign}
\Lambda_1^0(0,0) = \frac{\mu^{2\veps_\gamma}\,e_r^2}{2}\,\gamma^0\,\int [\D^{D_e}p] [\D p_{E0}]\,\frac{p_{E0}^2 - v^2 |\vec{p}\,|^2}{[p_{E0}^2 + v^2 |\vec{p}\,|^2]^2\,|\vec{p}\,|} = 0\, ,
\end{flalign}
where the frequency integral vanishes identically, see App.~\ref{App:MI} for some useful integrals. 
The temporal part of the vertex is therefore not renormalized at one-loop.  Graphically, this result can be summarized as follows:
\be
\delta Z_{1\Gamma^0} = - \mathcal{K}\bigg[ \Lambda_1^0/\gamma^0 \bigg] ~ = ~ -
\mathcal{K}\bigg[~~
      \parbox{15mm}{
    \begin{fmfgraph*}(15,15)
      \fmfleft{in}
      \fmfright{e1,e2}
      \fmf{boson}{in,vi}
      \fmf{plain}{e1,v1}
      \fmf{plain,tension=0.7}{v1,vi}
      \fmf{plain,tension=0.7}{vi,v2}
      \fmf{plain}{v2,e2}
      \fmffreeze
      \fmf{boson,right,tension=0.7}{v1,v2}
      \fmfv{decor.shape=circle,decor.filled=empty,decor.size=2thick}{vi,v1,v2}
    \end{fmfgraph*}
}~~
\bigg] ~  = 0 \, .
\label{WI1}
\ee
Together with Eq.~(\ref{Zpsi}), this implies that: $Z_{1\psi} = Z_{1\Gamma^0} = 1$
and the Ward identity, $Z_{\psi} = Z_{\Gamma^0}$, is (trivially) satisfied at one-loop.

Let's now turn to the vector part of the vertex, see Fig.~\ref{fig:one-loop-vertex}c, focusing again of the case where $k=q=0$.
Performing the trace and going to euclidean space yields:
\begin{flalign}
\vec{\Lambda}_1(0,0) = \frac{\mu^{2\veps_\gamma}\,e_r^2}{2}\,\vec{\gamma}\,\int [\D^{D_e}p] [\D p_{E0}]\,\frac{p_{E0}^2 + v^2 |\vec{p}\,|^2 (D_e-2)/D_e}{[p_{E0}^2 + v^2 |\vec{p}\,|^2]^2\,|\vec{p}\,|}\, ,
\end{flalign}
where now the frequency integral is non-zero. Performing the remaining integrations, we arrive at:
\be
\vec{\Lambda}_1(0,0) = \vec{\gamma}\,\,\frac{\al_g}{8}\,\left(\frac{\overline{\mu}^{\,2}}{m^2}\right)^{\veps_\gamma}\,\frac{e^{\gamma_E\veps_\gamma}\,\Gamma(1+\veps_\gamma)}{\veps_\gamma}\, ,
\ee
which shows that the vector part of the vertex is UV singular ($m$ is just an arbitrary IR regulator). Extracting the pole part, the corresponding renormalization constant together with its graphical representation
are given by:
\be
\delta Z_{1 \vec{\Gamma}\,} = - \mathcal{K}\bigg[ \vec{\Lambda}_1/\vec{\gamma}\, \bigg] ~ = ~ -
\mathcal{K}\bigg[~~
      \parbox{15mm}{
    \begin{fmfgraph*}(15,15)
      \fmfleft{in}
      \fmfright{e1,e2}
      \fmf{boson}{in,vi}
      \fmf{plain}{e1,v1}
      \fmf{plain,tension=0.7}{v1,vi}
      \fmf{plain,tension=0.7}{vi,v2}
      \fmf{plain}{v2,e2}
      \fmffreeze
      \fmf{boson,right,tension=0.7}{v1,v2}
      \fmfv{decor.shape=circle,decor.filled=shaded,decor.size=2thick}{vi}
      \fmfv{decor.shape=circle,decor.filled=empty,decor.size=2thick}{v1,v2}
    \end{fmfgraph*}
}~~
\bigg] ~  = - \frac{\al_{gr}}{8 \veps_\gamma} \, .
\label{vecLambda1}
\ee
At one-loop, this result is consistent with the Ward identity: $Z_{\vec{\Gamma}\,} = Z_{v} Z_{\psi}$ which may be graphically represented as:
\be
\mathcal{K}\bigg[~~
      \parbox{15mm}{
    \begin{fmfgraph*}(15,15)
      \fmfleft{in}
      \fmfright{out}
      \fmf{plain}{in,vi}
      \fmf{plain,tension=0.2}{vi,vo}
      \fmf{boson,left,tension=0.2}{vi,vo}
      \fmf{plain}{vo,out}
      \fmfv{decor.shape=circle,decor.filled=empty,decor.size=2thick}{vi,vo}
    \end{fmfgraph*}
}~~
\bigg] ~ = ~ -
\mathcal{K}\bigg[~~
      \parbox{15mm}{
    \begin{fmfgraph*}(15,15)
      \fmfleft{in}
      \fmfright{e1,e2}
      \fmf{boson}{in,vi}
      \fmf{plain}{e1,v1}
      \fmf{plain,tension=0.7}{v1,vi}
      \fmf{plain,tension=0.7}{vi,v2}
      \fmf{plain}{v2,e2}
      \fmffreeze
      \fmf{boson,right,tension=0.7}{v1,v2}
      \fmfv{decor.shape=circle,decor.filled=shaded,decor.size=2thick}{vi}
      \fmfv{decor.shape=circle,decor.filled=empty,decor.size=2thick}{v1,v2}
    \end{fmfgraph*}
}~~
\bigg] ~\, .
\label{WI2}
\ee
Notice that the peculiar Ward identities Eqs.~(\ref{WI1}) and (\ref{WI2}) are rather unusual with respect to those which can be found in the case of usual (Lorentz-invariant) QEDs.
This will play a crucial role in deriving the correct interaction correction to the optical conductivity in the non-relativistic limit.

\section{Optical conductivity at two loops} 
\label{Sec:TwoLoop}

\begin{figure}
        \includegraphics{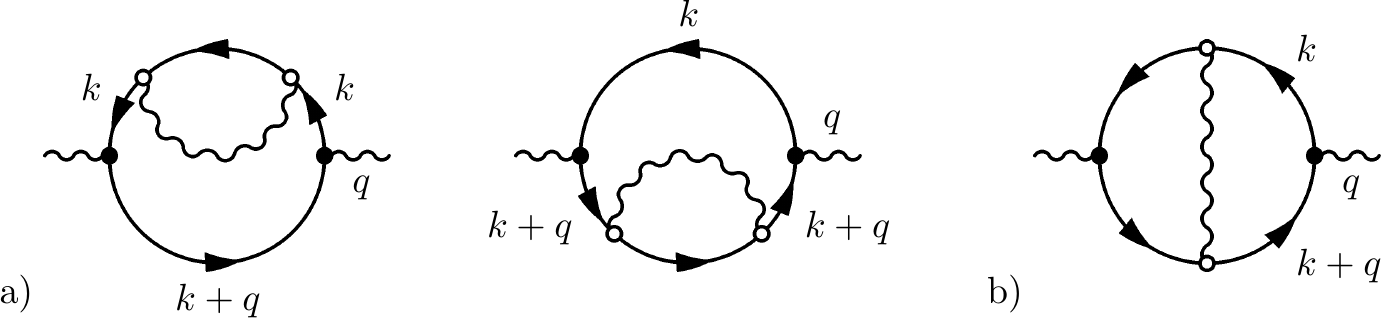}
    \caption{\label{fig:2loop-polarization-cc}
     Two-loop vacuum polarization, $\Pi_2^{\mu \nu}$, diagrams.}
\end{figure}
%

\begin{figure}
        \includegraphics{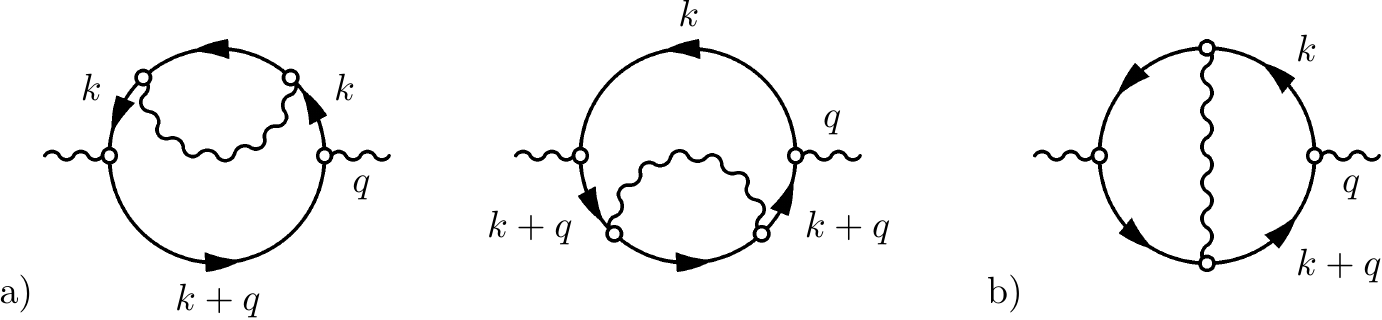}
    \caption{\label{fig:2loop-polarization-dd}
     Two-loop vacuum polarization, $\Pi_2^{0 0}$, diagrams.}
\end{figure}
%

We now proceed on computing the 2-loop corrections displayed on Fig.~\ref{fig:2loop-polarization-cc}:
$\Pi_2^{\mu \nu}(q) = 2 \Pi_{2a}^{\mu \nu}(q) + \Pi_{2b}^{\mu \nu}(q)$
where $\Pi_{2a}$ is the so-called self-energy correction and $\Pi_{2b}$ is the so-called vertex correction. The latter are defined in the usual way as:
\begin{subequations}
\label{def:pi2loop}
\begin{flalign}
&\I \Pi_{2a}^{\mu \nu}(q) =
- \int [\D^{d_e}k ]\, \Tr \left[ (-\I e\gamma^\nu) \,S_0(k+q)\,(-\I e\gamma^\mu)\, S_0(k)\left( -\I \Sigma_1(k) \right)\,S_0(k) \right]\, ,
\label{def:pi2a} \\
&\I \Pi_{2b}^{\mu \nu}(q) = - \int [\D^{d_e}k]\,\Tr \left[ (-\I e\gamma^\nu)\, S_0(k+q)\,(-\I e \Lambda_1^\mu(k,q))\, S_0(k) \right]\, ,
\label{def:pi2b}
\end{flalign}
\end{subequations}
where the fermion self-energy was defined in Eq.~(\ref{def:fermion-self}) and the fermion-photon vertex in Eq.~(\ref{def:vertex}).
For completeness, the diagrams associated to $\Pi^{00}(a)$ are displayed on Fig.~\ref{fig:2loop-polarization-dd}. Our convention for the conductivity will be that:
\begin{subequations}
\label{def:sigma2dd}
\begin{flalign}
	&\sigma_2 = \sigma_{2a} + \sigma_{2b}\, , 
	\\
	&\sigma_{2a} = - \lim_{\vec{q} \ra 0} \, \frac{\I q_0}{|\vec{q}\,|^2}\,2 \,\Pi_{2a}^{00}(q_0,\vec{q}\,)\, , 
	\\
	&\sigma_{2b} = - \lim_{\vec{q} \ra 0} \, \frac{\I q_0}{|\vec{q}\,|^2}\,\Pi_{2b}^{00}(q_0,\vec{q}\,)\, ,
\end{flalign}
\end{subequations}
and similarly for $\tilde{\sigma}_2$.

\subsection{Density-density correlation function approach}

The computation of the self-energy diagrams displayed on Fig.~\ref{fig:2loop-polarization-dd} has been performed
in Ref.~\cite{Teber:2014ita} (see also App.~B). We briefly recall the final results for clarity:
\begin{subequations}
\label{pi200}
\begin{flalign}
&2\,\Pi_{2a}^{00}(q_0,\vec{q} \ra 0) = - \frac{N_F\, e_r^2}{8}\,\frac{|\vec{q}\,|^2}{\I q_0}\,\frac{\al_{gr}}{2}\, ,
\label{pi2a00} \\
	&\Pi_{2b}^{00}(q_{0},\vec{q} \ra 0) = -\frac{N_F \,e_r^2}{8}\,\frac{|\vec{q}\,|^2}{\I q_0}\,\al_{gr}\,\frac{8-3 \pi}{6}\, ,
\label{pi2b00}
\end{flalign}
\end{subequations}
where, with two-loop accuracy, the coupling constant is the renormalized one.
Combining these results with Eqs.~(\ref{def:sigma2dd}) yields:
\begin{subequations}
\label{sigma2dd}
\begin{flalign}
	&\sigma_{2a}(q_0) = \frac{N_F\, e_r^2}{8}\,\frac{\al_{gr}}{2}\, ,
\label{sigma2a} \\
	&\sigma_{2b}(q_0) = \frac{N_F\, e_r^2}{8}\,\al_{gr}\,\frac{8-3 \pi}{6}\, ,
\label{sigma2b}
\end{flalign}
\end{subequations}
from which the total (bare) optical conductivity reads:
\be
\sigma_{2}(q_0) = \frac{N_F\, e_r^2}{8}\,\al_{gr}\,\frac{11-3 \pi}{6}\, ,
\ee
with a (bare) interaction correction coefficient corresponding to $\mathcal{C}^{(3)}$.

We now proceed on computing the renormalized conductivity using the BPHZ prescription. According to the latter, the renormalized diagrams contributing to the density-density 
correlation function, $\Pi_{2\,\al\,r}^{00}(q)$ ($\al = a,b$), are related to the bare ones as follows:
\begin{subequations}
\label{BPHZ:dd}
\begin{flalign}
	&\Pi_{2\,\al\,r}^{00} = \mathcal{R}\,\Pi_{2\,\al}^{00} = (1 - \mathcal{K})\,\mathcal{R}'\,\Pi_{2\,\al}^{00} \quad (\al=a,\,b)\, ,
	\\
	&\mathcal{R}'\,\Pi_{2\,\al}^{00} = \Pi_{2\al}^{00} + \Pi_{2\al'}^{00}\, ,
\end{flalign}
\end{subequations}
where $\mathcal{R}'$ is the so-called incomplete $R$-operation because it subtracts only the subdivergences and $\Pi_{2\al'}^{00}$ may be graphically represented as:
\begin{subequations}
\label{R':a+b}
\begin{flalign}
	&\Pi_{2a'}^{00} = -
\mathcal{K}\, \bigg[~
   \parbox{15mm}{
    \begin{fmfgraph*}(15,15)
      \fmfleft{in}
      \fmfright{out}
      \fmf{plain}{in,vi}
      \fmf{plain,tension=0.2}{vi,vo}
      \fmf{boson,left,tension=0.2}{vi,vo}
      \fmf{plain}{vo,out}
      \fmfv{decor.shape=circle,decor.filled=empty,decor.size=2thick}{vi,vo}
    \end{fmfgraph*}
}~ \bigg]~\star~
\parbox{15mm}{
    \begin{fmfgraph*}(15,15)
      \fmfleft{in}
      \fmfright{out}
      \fmf{boson}{in,ve}
      \fmf{plain,right,tension=0.2}{ve,vw}
      \fmf{plain,right,tension=0.2}{vw,ve}
      \fmf{boson}{vw,out}
      \fmfv{decor.shape=circle,decor.filled=empty,decor.size=2thick}{ve,vw}
    \end{fmfgraph*}
}\, ~~, 
\label{R':a} \\
	&\Pi_{2b'}^{00} = -2
\mathcal{K}\, \bigg[~
   \parbox{15mm}{
    \begin{fmfgraph*}(15,15)
      \fmfleft{in}
      \fmfright{e1,e2}
      \fmf{boson}{in,vi}
      \fmf{plain}{e1,v1}
      \fmf{plain,tension=0.7}{v1,vi}
      \fmf{plain,tension=0.7}{vi,v2}
      \fmf{plain}{v2,e2}
      \fmffreeze
      \fmf{boson,right,tension=0.7}{v1,v2}
      \fmfv{decor.shape=circle,decor.filled=empty,decor.size=2thick}{vi,v1,v2}
    \end{fmfgraph*}
}~ \bigg]~\star~
\parbox{15mm}{
    \begin{fmfgraph*}(15,15)
      \fmfleft{in}
      \fmfright{out}
      \fmf{boson}{in,ve}
      \fmf{plain,right,tension=0.2}{ve,vw}
      \fmf{plain,right,tension=0.2}{vw,ve}
      \fmf{boson}{vw,out}
      \fmfv{decor.shape=circle,decor.filled=empty,decor.size=2thick}{ve,vw}
    \end{fmfgraph*}
}\, ~~.
\label{R':b}
\end{flalign}
\end{subequations}
The peculiarity of the present non-relativistic theory is that the one-loop fermion self-energy and fermion-photon vertex subgraphs appearing in Eq.~(\ref{R':a+b}) are not related by a Ward identity
and therefore do not cancel each other (contrarily to what happens in usual QED). The case of $\Pi_{2b}^{00}$ is trivial: this diagram is finite overall so 
$\mathcal{K} \Pi_{2b}^{00}=0$ and, from Eq.~(\ref{WI1}), its subgraphs is also finite so: $\Pi_{2b'}^{00}=0$. This diagram is therefore absolutely convergent in 
Weinberg's sense~\cite{Weinberg:1959nj} so that:
\be
\sigma_{2br}(q_0) =  \sigma_{2b} = \frac{N_F\, e_r^2}{8}\,\al_{gr}\,\frac{8-3 \pi}{6}\, .
\label{sigma2br}
\ee
The case of $\Pi_{2a}^{00}$ is more interesting: this diagram is also finite overall, so: $\mathcal{K} \Pi_{2a}^{00}=0$. However, its subgraph is divergent, see Eq.~(\ref{dZ1v}),
and needs to be subtracted. In order to compute $\Pi_{2a'}^{00}$ we go to the integral representation of Eq.~(\ref{R':a}) as the $\star$ operation
does not reduce to a simple multiplication (the diagram is not logarithmic and care must be taken in projecting out its pseudo-Lorentz structure). 
Performing the trace, going to euclidean space, integrating over frequencies and taking the $\vec{q} \ra 0$ limit
leads to:
\begin{flalign}
\Pi_{2a'}^{00}(m,\vec{q} \ra 0) =  -\frac{N_F \, e_r^2}{4v}\,\mu^{2\veps_\gamma}\,|\vec{q}\,|^2\,\frac{D_e-1}{D_e}\,\int [\D^{D_e} k]\,\mathcal{K} \bigg[ \Sigma_{1k}(|\vec{k}^{\,2}|) \bigg]\,
\frac{|\vec{k}\,|^2 - m^2}{|\vec{k}\,|\, [|\vec{k}\,|^2 + m^2]^2}\, ,
\label{pi2a'-inter}
\end{flalign}
where $m = q_{E0}/2 v$. Substituting Eq.~(\ref{dZ1v}) and computing the integral using the formulas of App.~\ref{App:MI} yields:
\begin{flalign}
\Pi_{2a'}^{00}(m,\vec{q} \ra 0) = \frac{N_F e_r^2}{16\pi v}\,\frac{|\vec{q}\,|^2}{m}\,\frac{\al_{gr}}{8\veps_\gamma}\,\left(\frac{\overline{\mu}^{\,2}}{m^2}\right)^{\veps_\gamma}\,
\frac{(D_e-1)\,(D_e-2)}{D_e}\,e^{2\gamma_E\veps_\gamma}
\,B(1,1/2)\, ,
\end{flalign}
which is finite in the limit $\veps_\gamma \ra 0$ and reads:
\be
\Pi_{2a'}^{00}(q_0,\vec{q} \ra 0) = \frac{N_F e_r^2}{8}\,\frac{|\vec{q}\,|^2}{\I q_0}\,\frac{\al_{gr}}{8}\, .
\ee
The fact that both $\Pi_{2a'}^{00}$ and $\Pi_{2a}^{00}$ are finite implies that $\mathcal{K}\,\mathcal{R}'\,\Pi_{2\,a}^{00}=0$ which was expected since there is no global divergence to subtract.
However, the subtraction of the subdivergence brings a {\it finite} contribution to the renormalized function:
\bea
2\Pi_{2ar}^{00}(q_0,\vec{q} \ra 0) = 2\Pi_{2a}^{00} + 2\Pi_{2a'}^{00} 
= -\frac{N_F e_r^2}{8}\,\frac{|\vec{q}\,|^2}{\I q_0}\,\al_{gr}\,\biggl( \frac{1}{2} - \frac{1}{4} \biggr)\, .
\label{Summary}
\eea
The corresponding renormalized conductivity reads:
\be
\sigma_{2ar}(q_0) = \frac{N_F\, e_r^2}{8}\,\al_{gr}\,\biggl( \frac{1}{2} - \frac{1}{4} \biggr)\, ,
\label{sigma2ar}
\ee
and is decreased with respect to its bare value Eq.~(\ref{sigma2a}) by a quantity:  $\mathcal{C}'=-1/4$. 
Hence,  the total two-loop renormalized conductivity reads:
\bea 
\sigma_{2r}(q_0) = \sigma_{2ar} + \sigma_{2br} = \frac{N_F\, e_r^2}{8}\, \al_{gr}\,\biggl( \frac{11-3 \pi}{6} - \frac{1}{4} \biggr) 
= \frac{N_F\, e_r^2}{8}\, \al_{gr}\,\frac{19-6\pi}{12} \, , 
\label{sigma2r:final}
\eea
in accordance with the advertised result for the interaction correction coefficient, Eq.~(\ref{c-DR}).

\subsection{Current-current correlation function approach}

We now proceed in a similar way with the self-energy diagrams displayed on Fig.~\ref{fig:2loop-polarization-cc}. The latter
were computed in Ref.~\cite{Teber:2014ita} (see also App.~B). We briefly recall the final results for clarity:
\begin{subequations}
\label{pi2mumu-expr}
\begin{flalign}
	&2K_{2a}(q_{0}) = \frac{N_F\,e_r^2}{8\I q_0}\,\frac{\al_{gr}}{4}\,\left( -\frac{1}{\veps_\gamma} + 3 + 2 L_{q_0} - 4 \log 2 + \Ord(\veps_\gamma) \right)\, ,
\label{pi2amumu-expr} \\
&K_{2b}(q_0) = -2K_{2a}(q_0)
	+\frac{N_F\,e_r^2}{8\,\I q_{0}}\,\al_{gr}\,\frac{11 - 3\pi}{6}\, ,
\label{pi2bmumu-expr}
\end{flalign}
\end{subequations}
where again the coupling constant can be replaced by the renormalized one with two-loop accuracy and $L_{q_0}=\log(-q_0^2/(4v^2 \overline{\mu}^{\,2}))$. Notice
that the diagrams are now individually divergent contrarily to those of Eqs.~(\ref{pi200}).
Singular terms cancel from the sum of Eqs.~(\ref{pi2amumu-expr}) and (\ref{pi2bmumu-expr}). Combining these equations with Eq.~(\ref{sigma-cc2})
yields the total (bare) optical conductivity at two-loops: 
\be
\tilde{\sigma}_{2}(q_0) = \frac{N_Fe_r^2}{8}\,\al_{gr}\,\frac{11-3\pi}{6}\, ,
\ee
with a (bare) interaction correction coefficient corresponding once again to $\mathcal{C}^{(3)}$.

We now proceed on computing the renormalized conductivity using the BPHZ prescription.
Similarly to the density-density case, the renormalized diagrams contributing to the current-current correlation function, $K_{2\,\al\,r}(q_0)$ ($\al = a,b$), are related to the bare ones as follows:
\begin{subequations}
\label{BPHZ:cc}
\begin{flalign}
	&K_{2\,\al\,r} = \mathcal{R}\,K_{2\,\al} = (1 - \mathcal{K})\,\mathcal{R}'\,K_{2\,\al} \quad (\al=a,\,b)\, , 
	\\
	&\mathcal{R}'\,K_{2\,\al} = K_{2\al} + K_{2\al'}\, ,
\end{flalign}
\end{subequations}
where $K_{2\al'}$ may be graphically represented as:
\begin{subequations}
\label{R':Ka+b}
\begin{flalign}
&K_{2a'} = -
\mathcal{K}\, \bigg[~
   \parbox{15mm}{
    \begin{fmfgraph*}(15,15)
      \fmfleft{in}
      \fmfright{out}
      \fmf{plain}{in,vi}
      \fmf{plain,tension=0.2}{vi,vo}
      \fmf{boson,left,tension=0.2}{vi,vo}
      \fmf{plain}{vo,out}
      \fmfv{decor.shape=circle,decor.filled=empty,decor.size=2thick}{vi,vo}
    \end{fmfgraph*}
}~ \bigg]~\star~
\parbox{15mm}{
    \begin{fmfgraph*}(15,15)
      \fmfleft{in}
      \fmfright{out}
      \fmf{boson}{in,ve}
      \fmf{plain,right,tension=0.2}{ve,vw}
      \fmf{plain,right,tension=0.2}{vw,ve}
      \fmf{boson}{vw,out}
      \fmfv{decor.shape=circle,decor.filled=full,decor.size=2thick}{ve,vw}
    \end{fmfgraph*}
}\, , 
\label{R':Ka}	
\\
	&K_{2b'} = -2
\mathcal{K}\, \bigg[~
   \parbox{15mm}{
    \begin{fmfgraph*}(15,15)
      \fmfleft{in}
      \fmfright{e1,e2}
      \fmf{boson}{in,vi}
      \fmf{plain}{e1,v1}
      \fmf{plain,tension=0.7}{v1,vi}
      \fmf{plain,tension=0.7}{vi,v2}
      \fmf{plain}{v2,e2}
      \fmffreeze
      \fmf{boson,right,tension=0.7}{v1,v2}
      \fmfv{decor.shape=circle,decor.filled=full,decor.size=2thick}{vi}
      \fmfv{decor.shape=circle,decor.filled=empty,decor.size=2thick}{v1,v2}
    \end{fmfgraph*}
}~ \bigg]~\star~
\parbox{15mm}{
    \begin{fmfgraph*}(15,15)
      \fmfleft{in}
      \fmfright{out}
      \fmf{boson}{in,ve}
      \fmf{plain,right,tension=0.2}{ve,vw}
      \fmf{plain,right,tension=0.2}{vw,ve}
      \fmf{boson}{vw,out}
      \fmfv{decor.shape=circle,decor.filled=full,decor.size=2thick}{ve,vw}
    \end{fmfgraph*}
}\, ~~.
\label{R':Kb}
\end{flalign}
\end{subequations}
Contrarily to the density-density case, both $K_{2a}$ and $K_{2b}$ need to be properly renormalized. Indeed, in the present case, both one-loop fermion self-energy and fermion-photon vertex 
subgraphs appearing in Eqs.~(\ref{R':Ka+b}) are singular. One may wonder at this point if these contributions will cancel each other due to the Ward identity Eq.~(\ref{WI2}).
As will be shown in the following, this is not the case. The proof requires some care as the $\star$ operation does not (necessarily) reduce to a simple multiplication. 
Notice for example that only the vector part of the fermion-photon vertex subgraph is singular; but the full vertex appears in Eq.~(\ref{R':Kb}).
So, one needs to be careful in projecting out only the vector component. In order to do this we will use the integral representation of Eqs.~(\ref{R':Ka+b})

We first consider the integral representation of $K_{2a'}$. Performing the trace, going to euclidean space, integrating over frequencies and taking the $\vec{q} \ra 0$ limit
leads to:
\begin{flalign}
	K_{2a'}(m) = \frac{N_F e_r^2}{4v m^2}\,\frac{D_e-1}{D_e}\,\mu^{2\veps_\gamma}\,\int [\D^{D_e}k]\,\mathcal{K} \biggl[ \Sigma_{1k}(|\vec{k}^{\,2}|) \biggr]\,
\frac{|\vec{k}\,|\,(|\vec{k}\,|^2 - m^2)}{[|\vec{k}\,|^2 + m^2]^2}\, .
\label{pi2a'mumu-inter}
\end{flalign}
Substituting Eq.~(\ref{dZ1v}) and computing the integral yields:
\begin{flalign}
	K_{2a'}(m) = \frac{N_F e_r^2}{16\pi v m}\,\frac{\al_{gr}}{8 \veps_\gamma}\,\left(\frac{\overline{\mu}^{\,2}}{m^2}\right)^{2\veps_\gamma}\,(D_e-1)\,e^{\gamma_E\veps_\gamma}\,B(1,-1/2)\, .
\end{flalign}
The $\veps_\gamma$-expansion reads:
\be
2K_{2a'}(q_{0}) = \frac{N_F\,e_r^2}{8\I q_0}\,\frac{\al_{gr}}{4}\,\left( \frac{1}{\veps_\gamma} - 2 - L_{q_0} + \Ord(\veps_\gamma) \right)\, .
\label{pi2a'mumu-expr}
\ee
Adding the contribution of $2K_{2a}$ the pole terms cancel each other and we find that:
\be
2 K_{2\,ar} = 2 \mathcal{R}'\,K_{2\,a} = \frac{N_F\,e_r^2}{8\I q_0}\,\frac{\al_{gr}}{4}\,\biggl(1 + L_{q_0} - 4 \log 2 + \Ord(\veps_\gamma) \biggr)\, .
\label{K2ar}
\ee
The fact that $\mathcal{K} \mathcal{R}'\,K_{2\,a} = 0$ means that the overall counter-term is zero in accordance with the fact that the only singularity which needs to be subtracted is the
one associated with the fermion self-energy subgraph. From Eq.~(\ref{K2ar}), we may now derive the expression for the renormalized conductivity associated with 
Fig.~\ref{fig:2loop-polarization-cc}a:
\be
\tilde{\sigma}_{2ar}(q_0) = \frac{N_Fe_r^2}{8}\,\frac{\al_{gr}}{4}\,\biggl(1 + L_{q_0} - 4 \log 2 + \Ord(\veps_\gamma) \biggr)\, .
\label{tsigma2ar}
\ee

We now consider the case of  $K_{2b'}$. In order to compute the latter, we find it convenient to go back to the initial definition Eq.~(\ref{def:pi2b}) replacing
$\Lambda_1^\mu$ by $\vec{\gamma}\,\mathcal{K}\bigg[ \vec{\Lambda}_1/\vec{\gamma}\, \bigg]$ with the appropriate $\gamma$-matrix contractions. Performing the trace, going to euclidean space, 
integrating over frequencies and taking the $\vec{q} \ra 0$ limit leads to:
\begin{flalign}
	K_{2b'}(m) = - \frac{N_F e_r^2}{v m^2}\,\frac{D_e-1}{D_e}\,\mu^{2\veps_\gamma}\,\mathcal{K}\bigg[ \vec{\Lambda}_1/\vec{\gamma}\bigg]\,\int [\D^{D_e}k]\,\frac{|\vec{k}\,|}{|\vec{k}\,|^2 + m^2}\, .
\label{def:pi2b'}
\end{flalign}
The integral is easily computed with the rules for evaluation semi-massive tadpoles, see App.~\ref{App:MI}, and the result reads:
\begin{flalign}
	K_{2b'}(m) = - \frac{N_F e_r^2}{4\pi v m}\,\mathcal{K}\bigg[ \vec{\Lambda}_1/\vec{\gamma}\, \bigg]\,\left(\frac{\overline{\mu}^{\,2}}{m^2}\right)^{\veps_\gamma}\,
	\frac{D_e-1}{D_e}\,e^{\gamma_E\veps_\gamma}\,B(1,-1/2)\, .
\label{expr:pi2b'}
\end{flalign}
Substituting Eq.~(\ref{vecLambda1}) and performing the $\veps_\gamma$-expansion then yields:
\be
K_{2b'}(q_0) = \frac{N_F e_r^2}{8 \I q_0}\,\frac{\al_{gr}}{4}\,\left( -\frac{1}{\veps_\gamma} + 1 + L_{q_0} + \Ord(\veps_\gamma) \right)\, .
\label{expand:pi2b'}
\ee
Adding the contribution of $K_{2b}$, the pole terms cancel each other and we find that:
\begin{flalign}
	K_{2\,br} = 2 \mathcal{R}'\,K_{2\,b} = \frac{N_F\,e_r^2}{8\,\I q_{0}}\,\al_{gr}\,\frac{11 - 3\pi}{6}
	+ \frac{N_F\,e_r^2}{8\I q_0}\,\frac{\al_{gr}}{4}\,\biggl(-2 - L_{q_0} + 4 \log 2 + \Ord(\veps_\gamma) \biggr)\, .
\label{K2br}
\end{flalign}
The fact that $\mathcal{K} \mathcal{R}'\,K_{2\,b} = 0$ means that the overall counter-term is zero in accordance with the fact that the only singularity which needs to be subtracted is the
one associated with the vector photon-fermion vertex subgraph. From Eq.~(\ref{K2br}), we may now derive the expression for the renormalized conductivity associated with
Fig.~\ref{fig:2loop-polarization-cc}b:
\begin{flalign}
	\tilde{\sigma}_{2br}(q_0) = \frac{N_F\,e_r^2}{8}\,\al_{gr}\,\frac{11 - 3\pi}{6} + \frac{N_Fe_r^2}{8}\,\frac{\al_{gr}}{4}\,\biggl(-2 - L_{q_0} + 4 \log 2 + \Ord(\veps_\gamma) \biggr)\, .
\label{tsigma2br}
\end{flalign}

Hence, adding Eqs.~(\ref{tsigma2ar}) and (\ref{tsigma2br}), the total two-loop renormalized conductivity reads:
\begin{flalign}
	\tilde{\sigma}_{2r}(q_0) = \tilde{\sigma}_{2ar} + \tilde{\sigma}_{2br} = \frac{N_F\, e_r^2}{8}\, \al_{gr}\,\biggl( \frac{11-3 \pi}{6} - \frac{1}{4} \biggr) 
	= \frac{N_F\, e_r^2}{8}\, \al_{gr}\,\frac{19-6\pi}{12} \, ,
\label{tsigma2r:final}
\end{flalign}
in accordance with the advertised result for the interaction correction coefficient, Eq.~(\ref{c-DR}), 
and therefore with the result obtained from the density-density correlation function, Eq.~(\ref{sigma2r:final}).

\section{Conclusion} 
\label{Sec:Conclusion}

In this paper, we have elaborated a field theoretic renormalization approach to the computation of electron-electron interactions in graphene and related planar Dirac 
liquids in the non-relativistic limit ($v/c \ra 0$) with instantaneous Coulomb interaction. As we have seen, the broken Lorentz invariance brings some peculiarities 
with respect to the relativistic case (non-standard Ward identities) as well as some complications (semi-massive Feynman diagrams instead of the massless
ones). Focusing on the optical conductivity of these materials, we have provided a clear proof that radiative corrections affecting this experimentally 
relevant observable are {\it finite} and {\it well determined}. Our proof makes use of the powerful BPHZ prescription. It confirms the validity of our previous approach 
based on conventional renormalization~\cite{Teber:2014ita} but is considerably more robust and allowed us to perform a refined diagram by diagram analysis.  
Both the density (where individual diagrams are overall finite) and Kubo formula (where individual diagrams are explicitly singular) 
approaches were shown to yield a single well-defined result for the two-loop interaction correction to the conductivity: $\mathcal{C}^{(2)} = (19-6\pi)/12 \approx 0.013$, 
a result first found by Mishchenko~\cite{Mishchenko2008} and which is compatible with experimental uncertainties~\cite{Nair:2008zz,PhysRevLett.101.196405}.
At this point, let's recall that Mishchenko's analysis~\cite{Mishchenko2008} warns against the use of the Kubo formula with a hard cut-off regulator. Reassuringly, 
in dimensional regularization no such problem is encountered and both density-density and current-current approaches can be safely used. Our formalism, can be  
extended to higher orders,~\footnote{As we have discussed in the Introduction, for coupling constant $\al_g \approx 1$, perturbation theory is highly questionable. 
For the polarization operator, the appearance of a small numerical constant $\mathcal{C}^{(2)}$ in factor of $\alpha_g$ brings a small parameter $\mathcal{C}^{(2)}\alpha_g$.  This may restore the validity of perturbation theory for the optical conductivity provided that higher 
order terms may be neglected as well. However, beyond first order, the value of $\mathcal{C}_n$ 
at order $n$ is unknown and it is therefore an open question as to whether or not $\mathcal{C}_n\alpha_g^n$ is small.} 
 other quantities and/or systems of other dimensionality, {\eg}, the optical conductivity of 3D Dirac materials.
It also constitutes a solid base on which the perturbation theory may be optimized in order to deal with the strong-coupling 
problem.~\footnote{The computation of NLO corrections in Ref.~\cite{PhysRevLett.113.105502} was approximate as an NLO diagram with two-loop polarization insertion was neglected (see lines below
Eq.~(9) in that paper). It turns out that this important diagram is the one proportional to $\mathcal{C}\alpha_g$. In another context, this kind of diagram was computed
exactly: see diagram 1 in Fig.~2 of Ref.~\cite{Kotikov:2016prf} together with Eqs.~(19) and (20) in this reference as well as the sentence below
Eq.~(19) in Ref.~\cite{Kotikov:2016yrn} where $\hat{\Pi}$ is a number analogous to $\mathcal{C}$. The fact that this number is small (as in Ref.~\cite{Kotikov:2016prf} 
where it is actually proportional to $\mathcal{C}^*$) or large (as in Ref.~\cite{Kotikov:2016yrn}) strongly impacts the results. Though this would require a more careful analysis,
our proof that $\mathcal{C} = \mathcal{C}^{(2)}$ is indeed small for graphene may further justify
the neglect of the corresponding diagram in Ref.~\cite{PhysRevLett.113.105502} which may in turn insure the good convergence of the RPA observed in Ref.~\cite{PhysRevLett.113.105502}.}
We leave these issues for our future investigations.

\acknowledgments

The work of A.V.K. was supported in part by the Russian Foundation for
Basic Research (Grant No.16-02-00790-a).


\appendix

\section{Master integrals}
\label{App:MI}

All integrals for counter-terms in the main text can be computed with the help of two basic integrals that we recall here.

The first one in the one-loop massless propagator-type integral with $n\leq 2$ which reads~\cite{Kazakov:1986mu}:
\begin{subequations}
\label{massless-p-integral}
\begin{flalign}
&\int [\D^{D} q]\,\frac{q^{\mu_1}\dotsc q^{\mu_n}
}{[q^2]^{\al}[(q-k)^2]^\beta}  =
\frac{(k^2)^{D/2-\al-\beta}}{(4\pi)^D}\,\left[k^{\mu_1} \dotsc k^{\mu_n}
\, G^{(n,0)}_0(\al,\beta)\, + \delta^2_n \frac{g^{\mu_1 \mu_2}}{D}
\, G^{(1,1)}_1(\al,\beta)\right], ~~~
\\
&G^{(n,m)}_i(\al,\beta) = \frac{a_n(\al)a_m(\beta)}{a_{n+m-i}
(\al+\beta-D/2-i)},
\\ 
&a_n(\al) = \frac{\Gamma(D/2-\al + n)}{\Gamma(\al)} \, ,
\end{flalign}
\end{subequations}
where $[\D^D q] = \D^{D} q / (2\pi)^{D}$ and  $\delta^2_n$ is the Kronecker
symbol. The simplified notation: $G(\al,\beta)=G^{(0,0)}_0(\al,\beta)$, will also be used.

The second one is the one-loop semi-massive tadpole diagram which reads:
\begin{subequations}
\label{tadpole}
\begin{flalign}
&\int \frac{[\D^D k]}{[k^2]^{\al}[k^2+m^2]^\beta} =
\frac{(m^2)^{D/2-\al-\beta}}{(4\pi)^{D/2}}\,B(\beta,\al)\, ,
\\ 
&B(\beta,\al) = \frac{\Gamma(D/2-\al)\,\Gamma(\al + \beta - D/2)}{\Gamma(D/2)\,\Gamma(\beta)}\, .
\end{flalign}
\end{subequations}
%



\section{Properties of two-loop self-energies}
\label{App:Two-loop}

Here we present some peculiar properties of the two-loop self-energies $\Pi_{2a}^{00}(q_0,\vec{q} \ra 0)$ and $K_{2a}(q_{0})$.

\subsection{$\Pi_{2a}^{00}(q_0,\vec{q} \ra 0)$}

Performing the trace, going to euclidean space, integrating over frequencies and taking the $\vec{q} \ra 0$ limit
leads to  \cite{Teber:2014ita}:
\begin{flalign}
\Pi_{2a}^{00}(m,\vec{q} \ra 0) =  \frac{N_F \, e_r^2}{4v}\,\mu^{2\veps_\gamma}\,|\vec{q}\,|^2\,\frac{D_e-1}{D_e}\,
\int [\D^{D_e} k]\,\Sigma_{1k}(|\vec{k}^{\,2}|) \,
\frac{|\vec{k}\,|^2 - m^2}{|\vec{k}\,|\, [|\vec{k}\,|^2 + m^2]^2}\, ,
\label{pi2a-inter1}
\end{flalign}
where, as in the main text, $m = q_{E0}/2 v$. Notice that this result can be represented in the following 
form:
\begin{flalign}
\Pi_{2a}^{00}(m,\vec{q} \ra 0) =  \frac{N_F \, e_r^2}{4v}\,\mu^{2\veps_\gamma}\,|\vec{q}\,|^2\,\frac{D_e-1}{D_e}\,\Sigma_{1k}(|\vec{k}^{\,2}|=1) \, J_{+}(\hat{a}=1) \, ,
\label{pi2a-J}
\end{flalign}
where
\begin{flalign}
&  J_{\pm}(\hat{a}) =
\int [\D^{D_e} k]\,
\frac{|\vec{k}\,|^2 - m^2}{[|\vec{k}\,|^2]^{\pm 1/2 + \hat{a} \veps_\gamma} \, [|\vec{k}\,|^2 + m^2]^2}\, .
\label{pi2a-inter2}
\end{flalign}
The one-loop self-energy $\Sigma_{1k}(|\vec{k}^{\,2}|)$ is singular at $\veps_\gamma\to0$ (see (\ref{fsigma2-k})), but the integral
 $J_{+}(\hat{a}) \sim O(\veps_\gamma)$ in this limit. Indeed:
\begin{flalign}
  J_{+}(\hat{a}) \sim B(2,-1/2 + \hat{a} \veps_\gamma) - B(2,1/2 + \hat{a} \veps_\gamma)\nonum 
= -2(\hat{a}+1) \veps_\gamma \,  B(1,1/2 + \hat{a} \veps_\gamma)\, ,
\label{J+}
\end{flalign}
with $B(\beta,\al)$ defined in Eq.~(\ref{tadpole}).

So, the result for $\Pi_{2a}^{00}(q_0,\vec{q} \ra 0)$ has the following form
\begin{flalign}
\Pi_{2a}^{00}(m,\vec{q} \ra 0) = \frac{N_F e_r^2}{16\pi v}\,\frac{|\vec{q}\,|^2}{m}\,\frac{\al_{gr}}{8}\,\left(\frac{\overline{\mu}^{\,2}}{m^2}\right)^{2\veps_\gamma}\,
	\frac{(D_e-1)\,2\veps_\gamma}{D_e}\,e^{2\gamma_E\veps_\gamma}\,G(1/2,1/2)\,B(1,1/2+\veps_\gamma)\, ,
\end{flalign}
which is finite in the limit $\veps_\gamma \ra 0$ and reads \cite{Teber:2014ita}:
\be
\Pi_{2a}^{00}(q_0,\vec{q} \ra 0) = \frac{N_F e_r^2}{16}\,\frac{|\vec{q}\,|^2}{v m}\,\frac{\al_{gr}}{8}\, .
\ee

The last integrations in $\Pi_{2a}^{00}(q_0,\vec{q} \ra 0)$ (see Eqs.~(\ref{pi2a-inter2}) and (\ref{pi2a-J})) and in
$\Pi_{2a'}^{00}(q_0,\vec{q} \ra 0)$ (see Eq.~(\ref{pi2a'-inter})) are related with the tadpoles $J_{+}(\hat{a}=1)$ and
$J_{+}(\hat{a}=0)$, respectively. Since at $\veps_\gamma \to 0$ $J_{+}(\hat{a}) \sim (\hat{a}+1) \veps_\gamma$ (see Eq.~(\ref{J+})),
then the counter-term $\Pi_{2a'}^{00}(q_0,\vec{q} \ra 0)$ equals half the value of $\Pi_{2a}^{00}(q_0,\vec{q} \ra 0)$ and thus the final
result of $\Pi_{2ar}^{00}(q_0,\vec{q} \ra 0)$ in (\ref{Summary}) is two times less than the one of $\Pi_{2a}^{00}(q_0,\vec{q} \ra 0)$.

\subsection{$K_{2a}(q_{0})$}

Performing the trace, going to euclidean space, integrating over frequencies and taking the $\vec{q} \ra 0$ limit
leads to \cite{Teber:2014ita}:
\begin{flalign}
K_{2a}(m) = -\frac{N_F e_r^2}{4v m^2}\,\frac{D_e-1}{D_e}\,\mu^{2\veps_\gamma}\,\int [\D^{D_e}k]\, \Sigma_{1k}(|\vec{k}^{\,2}|) \,
\frac{|\vec{k}\,|\,(|\vec{k}\,|^2 - m^2)}{[|\vec{k}\,|^2 + m^2]^2}\, .
\label{pi2a'mumu-interA}
\end{flalign}
As it was for $\Pi_{2a}^{00}(q_0,\vec{q} \ra 0)$,
this result can be represented in the following form:
\begin{flalign}
K_{2a}(m) =  - \frac{N_F \, e_r^2}{4v}\,\mu^{2\veps_\gamma}\,|\vec{q}\,|^2\,\frac{D_e-1}{D_e}\,\Sigma_{1k}(|\vec{k}^{\,2}|=1) \, J_{-}(\hat{a}=1) \, .
\label{pi2a-J-}
\end{flalign}
The one-loop self-energy $\Sigma_{1k}(|\vec{k}^{\,2}|)$ is singular at $\veps_\gamma \to 0$ (see (\ref{fsigma2-k})) and the integral
 $J_{-}(\hat{a})$ is finite in this limit:
\begin{flalign}
  J_{-}(\hat{a}) \sim B(2,-3/2 + \hat{a} \veps_\gamma) - B(2,-1/2 + \hat{a} \veps_\gamma)
= 2(1-(\hat{a}+1) \veps_\gamma) \,  B(1,-1/2 + \hat{a} \veps_\gamma)\, .
\label{J-}
\end{flalign}
So, the result for $K_{2a}(m)$ has the following form:
%
\begin{flalign}
K_{2a}(m) =  \frac{N_F e_r^2}{16\pi v m}\,\frac{\al_{gr}}{8}\,\left(\frac{\overline{\mu}^{\,2}}{m^2}\right)^{2\veps_\gamma}
\,\frac{2(D_e-1)^2}{D_e}\,\,e^{\gamma_E\veps_\gamma}\, \,G(1/2,1/2)\,B(1,-1/2+\veps_\gamma)\, ,
\end{flalign}
and its $\veps_\gamma$-expansion reads \cite{Teber:2014ita}:
\be
K_{2a}(q_{0}) = -\frac{N_F\,e_r^2}{16 vm}\,\frac{\al_{gr}}{8}\,\left( \frac{1}{\veps_\gamma} - 3 +4\ln 2 -2 L_{q_0} + \Ord(\veps_\gamma) \right)\, .
\label{pi2a'mumu-exprA}
\ee

\end{fmffile}
\end{document}